\documentclass{article}

\usepackage{graphicx}

\def\ligne#1{\hbox to \hsize{#1}}
\def\PlacerEn#1 #2 #3 {\rlap{\kern#1\raise#2\hbox{#3}}}

\newtheorem{fig}{Figure}

\newtheorem{axiom}{Axiom}
\newtheorem{post}{Postulate}
\newtheorem{defn}{Definition}
\newtheorem{thm}{Theorem}

\newtheorem{prop}{Proposition}
\newtheorem{cor}{Corollary}

\setbox221=\hbox{\includegraphics{cercle_L20.eps}}

\def\encercle#1#2{\hbox{\raise-5pt\copy221\hskip#2#1}}

\title{Using Grossone to count the number of elements of infinite sets and
the connection with bijections}
\author{Maurice Margenstern\\
Laboratoire d'Informatique Th\'eorique et Appliqu\'ee, EA 3097,\\
        Universit\'e Paul Verlaine $-$ Metz, UFR-MIM, \\
        \^Ile du Saulcy, 57045 Metz Cedex, France\\
and CNRS, LORIA\\
{\it e-mail}: {\tt margens@univ-metz.fr}
}

\begin{document}
\maketitle

\begin{abstract}
   In this paper, we look at how to count the number of elements of a set within
the frame of Sergeyev's numeral system. We also look at the connection between the
number of elements of a set and the notion of bijection in this new setting.
We also show the difference between this new numeral system and the results of the
traditional naive set theory.
\end{abstract}

\def\cqfd{\hbox{\kern 2pt\vrule height 6pt depth 2pt width 8pt\kern 1pt}}
\def\Hii{$I\!\!H^2$}
\def\Hiii{$I\!\!H^3$}
\def\Hiv{$I\!\!H^4$}
\def\norm{\hbox{$\vert\vert$}}
\def\grossone{\hbox{\textcircled{\bf 1}}}
\section{{\Large Introduction}}
\label{intro}
   This paper looks at a possible axiomatic foundation for the use of
bijections in the new methodology
introduced by Yaroslav {\sc Sergeyev} in his seminal papers, 
see~\cite{sergeyev1,sergeyev2,sergeyev3} which we will refer to as the
{\bf new numeral system}. This system contains the 
standard numeral system to write finite integers, 
positive and negatives. It also contains a symbol, 
\grossone, which is, by definition the number of elements
of the set of natural numbers with this property that
\hbox{$n<\grossone$} for any finite positive integer~$n$. 
We refer the reader to~\cite{sergeyev1,sergeyev3,sergeyevLL} for more
details and motivations on the system. 

   In Section~\ref{onbij}, we look again at the notion of bijection
in the traditional setting and, on an example, how it works in the
new setting of the new numeral system.

   In Section~\ref{yetbij}, we present a proposal toward a formalization
within the frame of the new numeral system.

   As we shall several times refer to the postulates of the new numeral system,
we reproduce them here for the convenience of the reader, exactly as they are
stated in~\cite{sergeyev3,sergeyevLL}.

\begin{post}
We postulate the existence of infinite and infinitesimal objects but accept that
human beings and machines are able to execute only a finite number of operations. 
\end{post}

\begin{post}
We shall not tell {\bf what are} the mathematical objects we deal with; we shall just 
construct more powerful tools that will allow us to improve our capacities to
observe and to describe properties of mathematical objects.
\end{post}

\begin{post}
We adopt the principle {\rm `The part is less than the whole'} to all numbers 
(finite, infinite and infinitesimal) and to all processes (finite and infinite).
\end{post}

\section{\Large Bijections and the principle `the part is less than the whole'}
\label{onbij}

   Remember that Cantor's set theory is based on the famous Bernstein theorem
which states the following assertion:

\begin{thm} $-$ {\rm (Bernstein)}\label{bernstein}
Let $A$ and~$B$ be sets such that there is an injective mapping~$f$ from~$A$ into~$B$
and an injective mapping~$g$ from~$B$ into~$A$. Then, there is a bijection~$\varphi$
from~$A$ onto~$B$.
\end{thm}
\def\reunion{\mathop{\cup}}
\def\linter{\mathop{\cap}}

   Traditionally, sets $A$ and~$B$ such that there is a bijection from~$A$ onto~$B$
are called {\bf equipotent}, we also usually say that they have the {\bf same number}
of elements. This relation between~$A$ and~$B$ is denoted by $A\equiv B$. If there is 
an {\bf injection} from~$A$ into~$B$, then  it is said that $A$ has no more elements
than~$B$ and this is denoted $A\leq B$. Bernstein's theorem says that this latter
relation defines an order among the sets. 

   Now, if we look at many examples of mathematical objects with the new tool 
given by the new numeral system, it seems that there is a blatant contradiction
between the just mentioned theorem and Postulate~3 of the new numeral system,
see~\cite{sergeyev3}. Indeed, in traditional mathematics, to have a part as big 
as the whole is the characteristics of the infinite sets. With Postulate~3,
this is no more true and a proper part of an infinite set is always less than
the whole. However, in his seminal papers, Yaroslav Sergeyev always stresses that
his new theory does not contradict Cantor's theory, that it simply gives new
tools to better study infinite objects than those provided by traditional set theory.

   We shall look at the following example which was the subject of a lightening
discussion with Yaroslav Sergeyev. The example is taken from geometry. Here it
is, as I presented it to Yaroslav.

   Let $A$ be a half-plane. Let $\delta$ be the line which is the border of~$A$,
see the left-hand side picture of Figure~\ref{planes}. Let $C$~be the reflection of~$A$
in~$\delta$, next picture of Figure~\ref{planes}.
Let $h$~be another line of the plane, parallel to~$\delta$ and inside~$A$
with $h\not=\delta$, third picture of Figure~\ref{planes}.
Let~$B$ be the reflection of~$C$ into~$h$, see the rightmost picture of
Figure~\ref{planes}. Clearly, $B\subset A$ and $B\not = A$. According
to Cantor's theory, $A$ and~$B$ have the same number of elements. According to
Sergeyev's system, as $B\subset A$ and $B\not = A$, $B$~has less elements than~$A$.
\vskip 10pt
\vtop{
\centerline{\hskip 10pt
\mbox{\includegraphics[width=33.3pt]{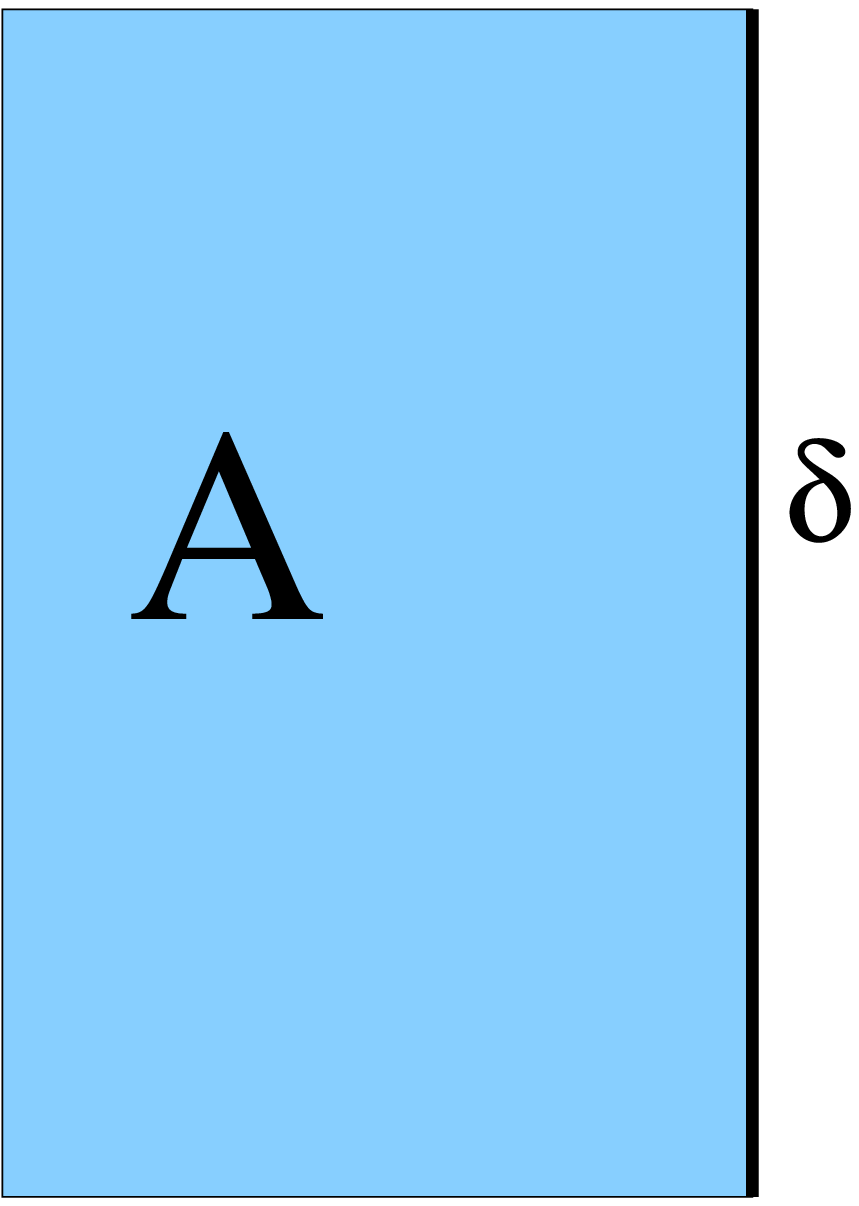}}
\hskip 40pt
\mbox{\includegraphics[width=60pt]{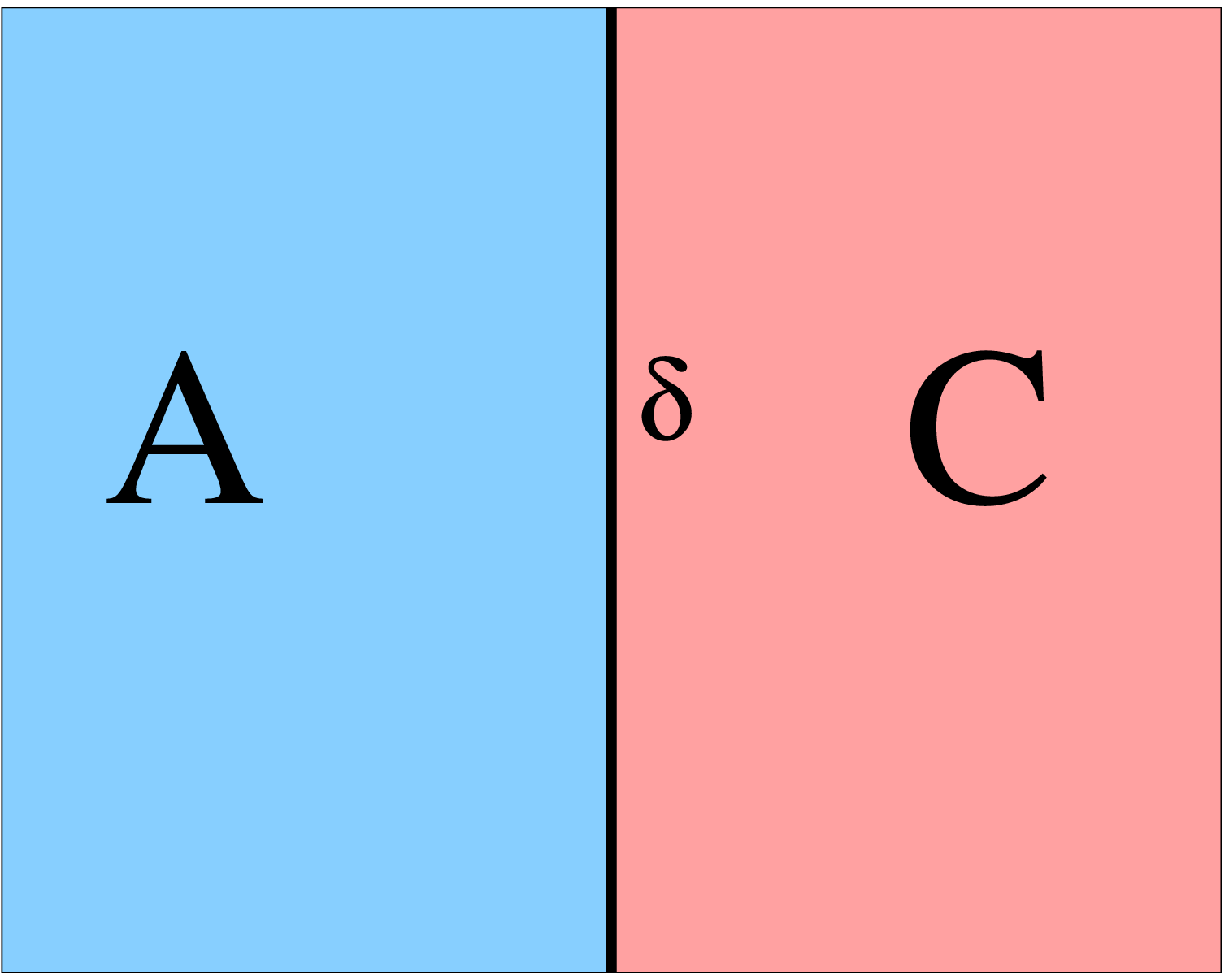}}
\hskip 10pt
\mbox{\includegraphics[width=60pt]{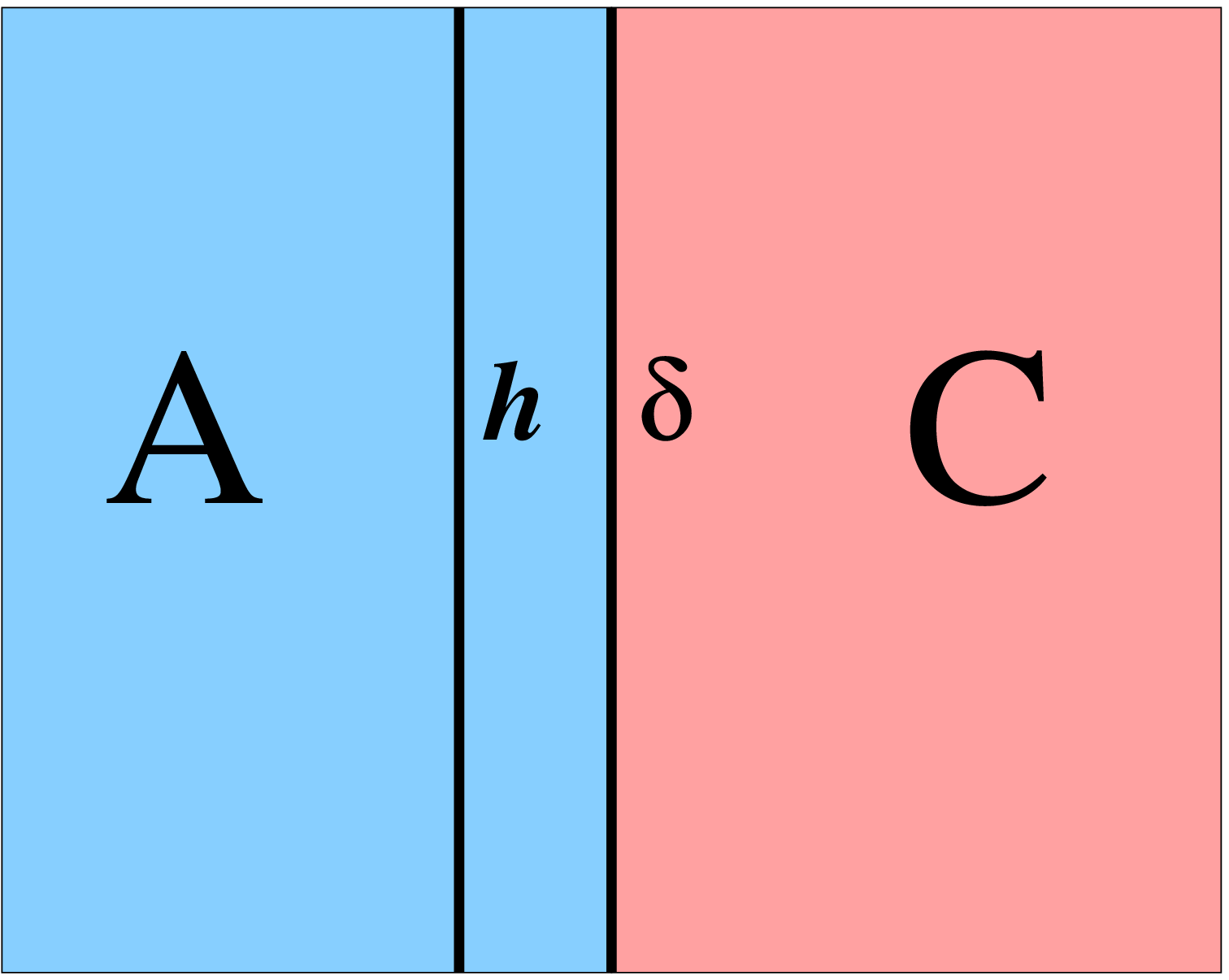}}
\hskip 10pt
\mbox{\includegraphics[width=60pt]{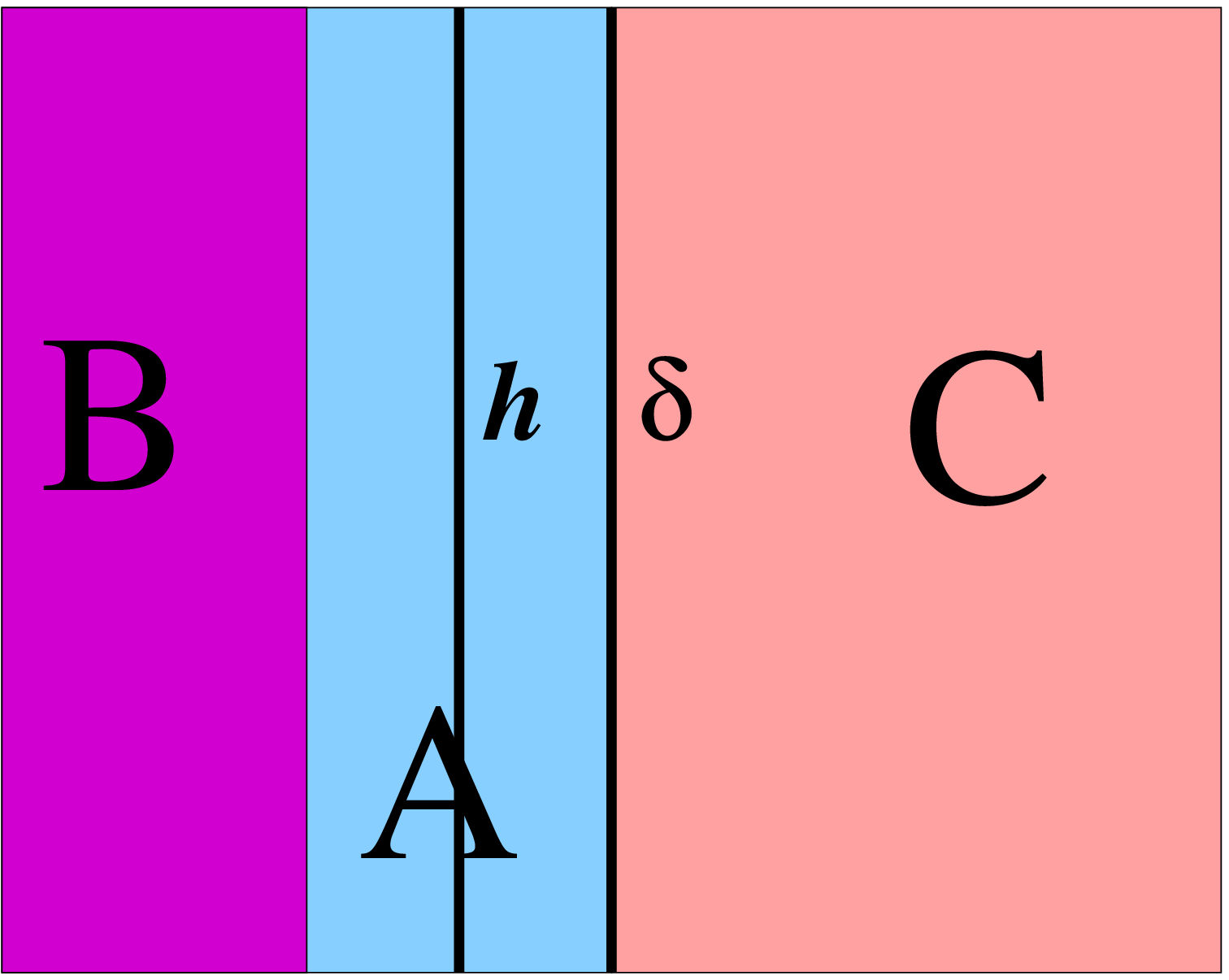}}
}
\begin{fig}\label{planes}
\small
An apparent contradiction between Cantor's theory and the new numeral system.
\end{fig}
}

   Yaroslav told me that something is not in agreement with his new approach
starting from the very presentation of the objects we consider. What is a half-plane?
What is a half-line as, in this example, the really objects at work are half-lines. 
Traditionally, we would write a half-line as $]-\infty,a]$ or $[a,+\infty[$
where $a$~is some real number. Yaroslav pointed that $-\infty$ and $+\infty$
are not precise notions. Compared with natural numbers, they are in the same 
position as {\it many} is with the natural numbers of 
Pirah$\tilde{\hbox{a}}$'s\footnote{In~\cite{sergeyev1,sergeyev2,sergeyev3},
Yaroslav mentions the discovery reported in~\cite{gordon} of an isolated group of 
people in Amazonia which have exactly the following numbers: $1$, $2$ and {\it many}.}, 
which are exactly 1 and~2, 
see~\cite{gordon,sergeyev1,sergeyev2,sergeyev3}. And so, 
Yaroslav continued,
we have to precise the bounds of the infinite interval of which consists our 
half-line: $[a,b]$ where $a$ and~$b$ are numbers, finite or infinite.
And then, he continued, let us make the computations associated with
the considered reflections.

   Here, I provide this computation, in order the reader could appreciate what
is found out. Let $A = [-b,a]\times I$, where $b$ is a positive infinite
number, $a$~is the abscissa of the point where $\delta$ cuts the $x$-axis 
and $I = [-c,c]$ is an infinite interval with $c$~a positive infinite
number. As our reflections are performed in axes which are perpendicular to
the $x$-axis, we perform the computations on abscissas only. The reflection 
in the line $\delta$ transforms $x$~into $-$$x$+$2a$. And so, we get
that $C = [a,b$+$2a]\times I$. Let $d$~be the abscissa of the points where the line~$h$
cuts the $x$-axis. Similarly, the reflection in~$h$ transforms $x$~into $-$$x$+$2d$,
so that $B = [$$-$$b$$-$$2a$+$2d,-a+2d]\times I$. Now, it is plain that 
$-$$a+2d<a$ as we assume
$d<a$ and that $-$$b$$-$$2a$+$2d< -$$b$, for the same reason. And this shows
us that $B\not\subset A$, contrary to what was concluded from the example,
see Figure~\ref{newplanes}.
Note that the same computations performed in the frame of Cantor's theory
shows that from $A = ]-\infty,a]\times L$ with $L = ]-\infty,+\infty[$,
we get $C = [a,+\infty[\times L$ and $B = ]-\infty,$$-$$a$+$2d]\times L$.
Accordingly, as Cantor's theory does not allow us to distinguish between
infinite quantities. We cannot see that the left-hand side bound of~$B$
is smaller than the left-hand side bound of~$A$ and so, there
are infinitely many points of~$B$ which are not contained in~$A$. 

\vskip 10pt
\vtop{
\centerline{\hskip 10pt
\mbox{\includegraphics[width=33.3pt]{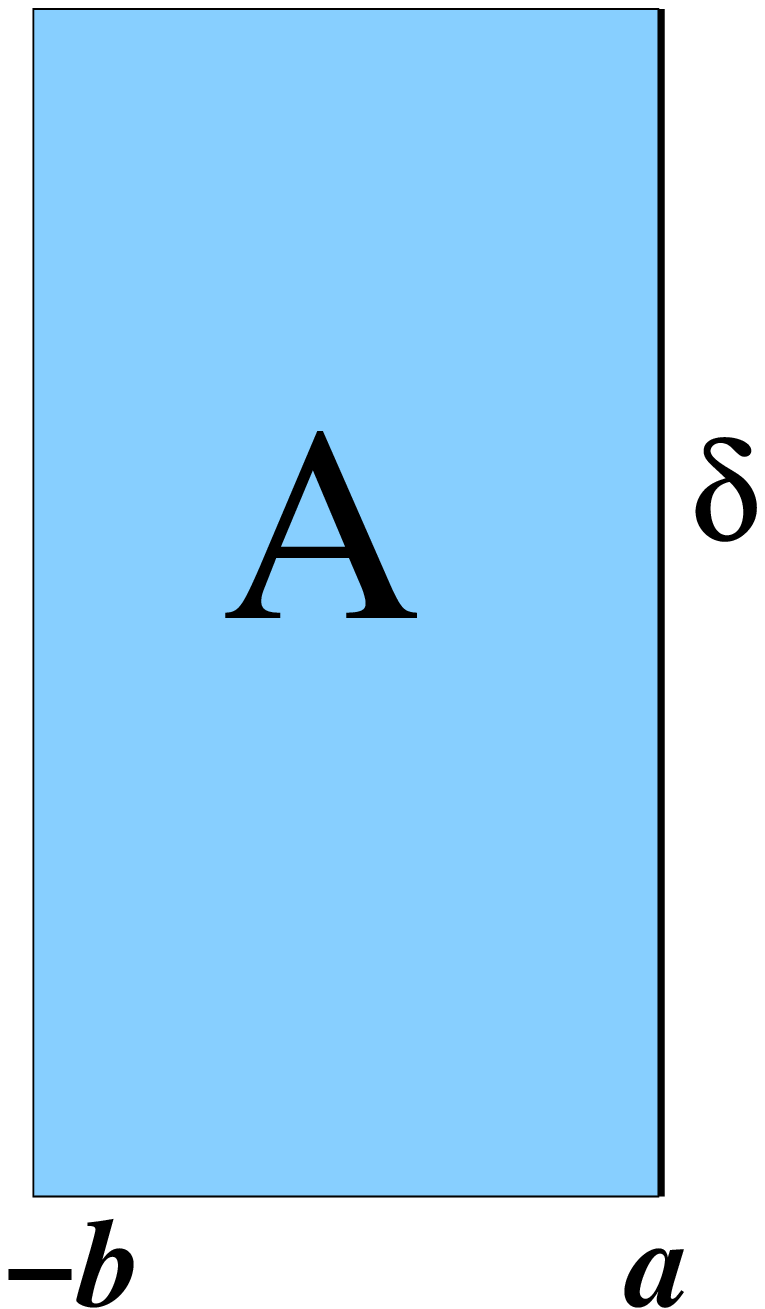}}
\hskip 20pt
\mbox{\includegraphics[width=60pt]{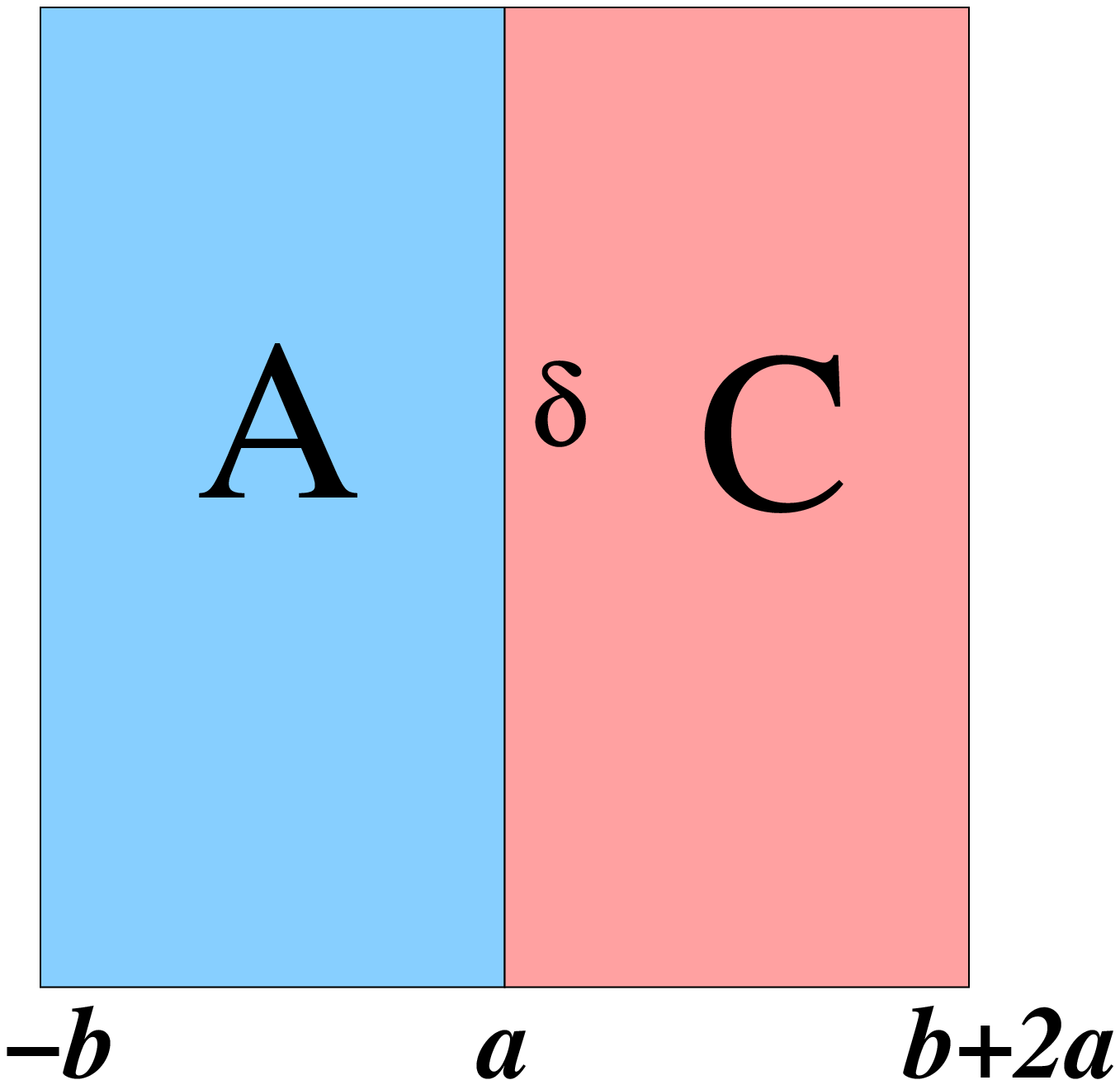}}
\hskip 10pt
\raise-5pt\hbox{\mbox{\includegraphics[width=60pt]{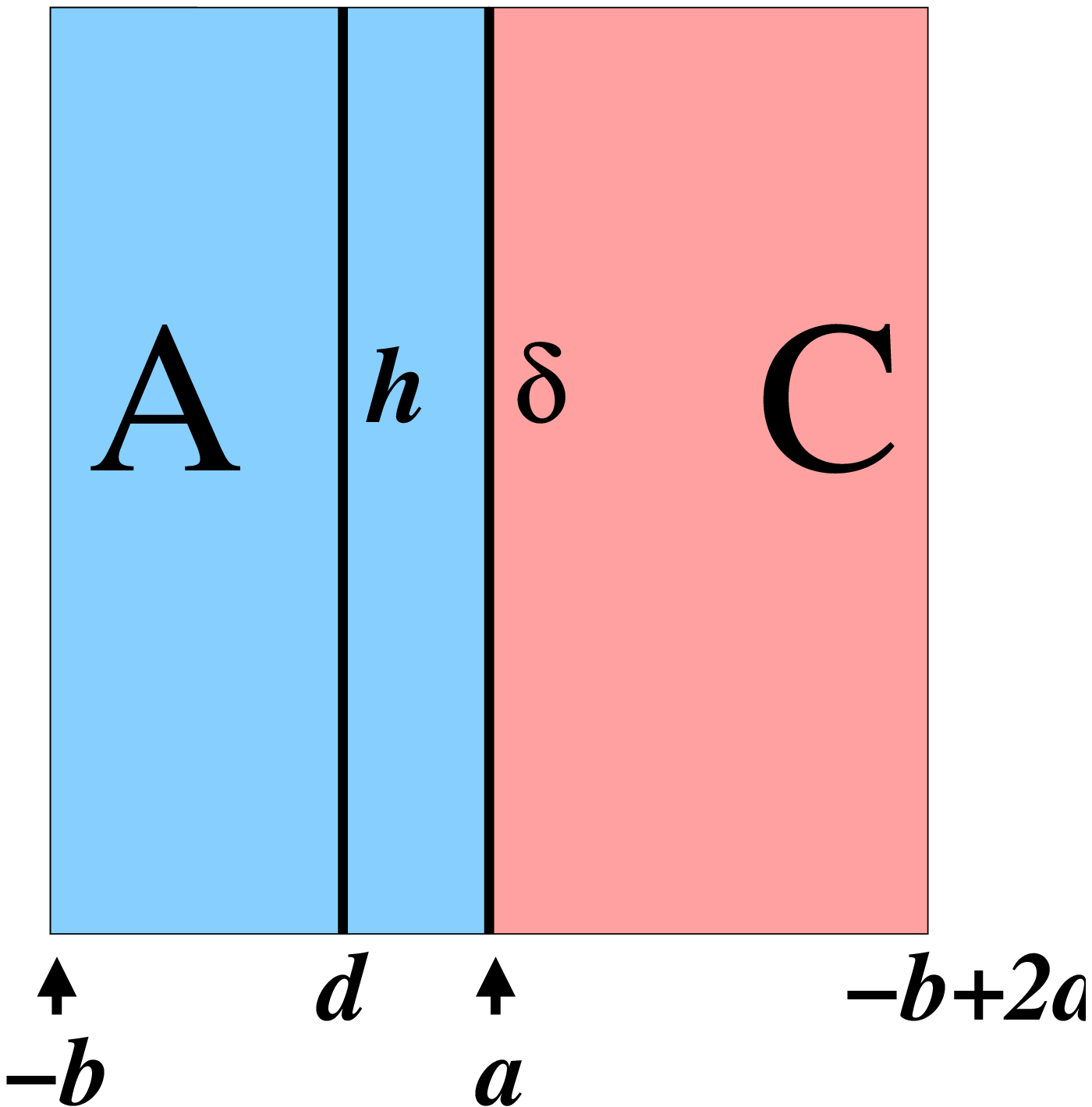}}}
\hskip 10pt
\raise-5pt\hbox{\mbox{\includegraphics[width=85pt]{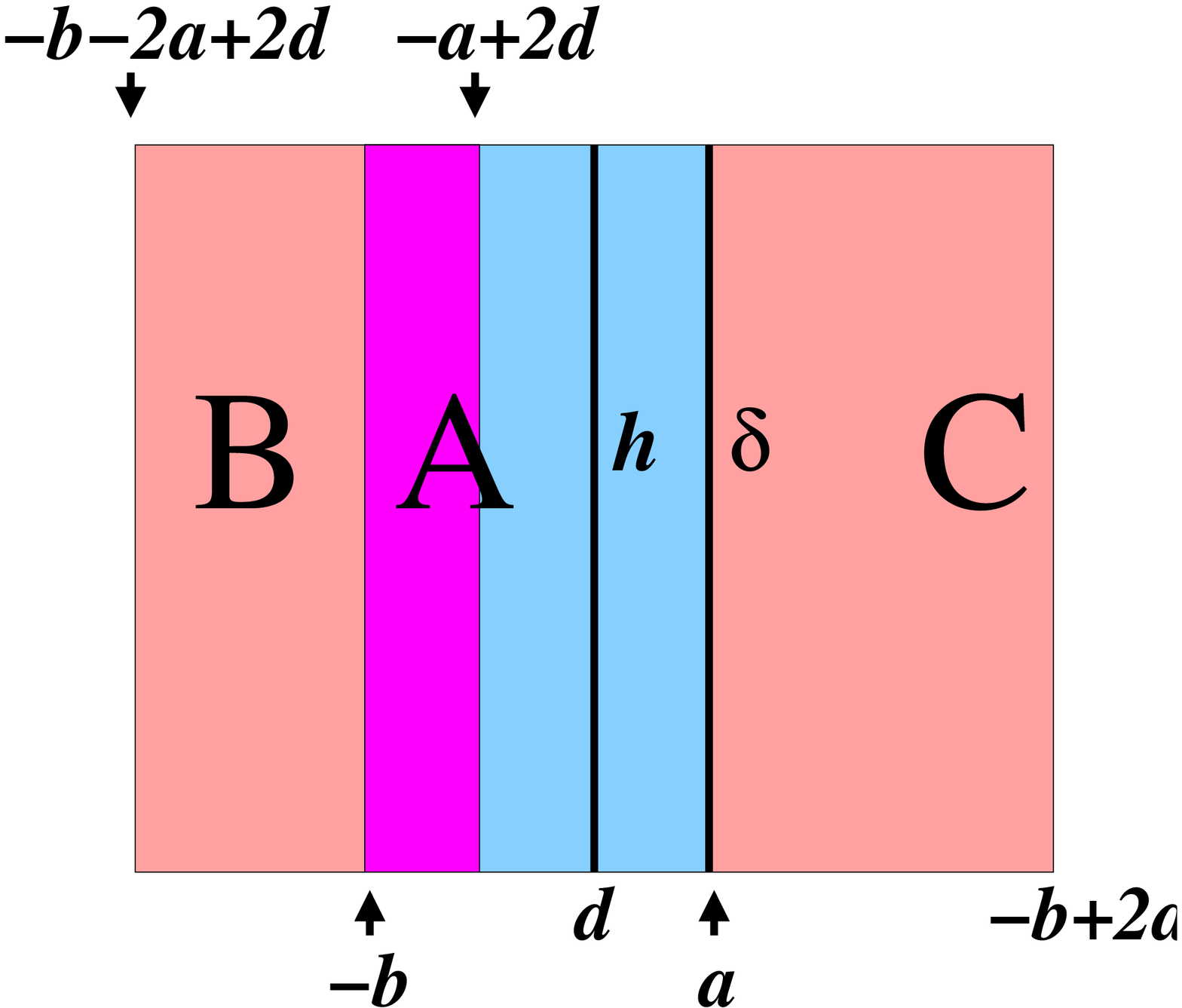}}}
}
\begin{fig}\label{newplanes}
\small
No contradiction: with the new numeral system, the definition of the images 
is more precise.
\end{fig}
}
   From this example, we can draw the conclusion: we have to make it as exact as 
we can the description of the objects on which we are working. In our example
this concerns the sets we considered and the operations which we performed
on these sets and, as far as numbers are involved, in which numeral system the numbers
are expressed. More precisely, we had to indicate the exact bounds of our
'half-planes', it should be better to call them semi-infinite strips,
{\it i.e. portions of the plane} in between by two parallel lines. We also
had to write down the transformations, here reflections in line, explicitly.

   It is important to notice that the new numeral system is not a sub-system
of non standard analysis and that it is neither a sub-system of the theory of
ordinals. These points will be made more clear a bit further.

\section{\Large How to use bijections in the new numeral system}
\label{yetbij}

   First, we look at the definition of the number of elements of a set
and then, how to deal with this notion in connection with bijections.

\subsection{The number of elements}
\label{subsec:enum}

   The simplest way to define the number of elements of a set~$E$ is to
count the number of its elements. Note that, practically, we can actually
count finite sets only, moreover, with a rather small number of elements.
This is contained in Postulate~1 of the new numeral system, see~\cite{sergeyev3}.

   Now, if we wish to perform abstract considerations, we have to bypass this possibility
and this is why we use descriptions. As long as we can describe an object, we may
consider that we handle it, if needed. Of course, what we can do for descriptions
strongly depends on the language we use to formulate them, this is the content
of Postulate 2, see~\cite{sergeyev3}.

   We have here to introduce a few notations which will be used in the statements and
the proofs which we provide later. In particular, if $\mu$ is a number of the numeral 
system, we denote by \hbox{$[1..\mu]$} the set of integers~$k$ which satisfy the relations
\hbox{$1\leq k$} and \hbox{$k\leq\mu$}. Here and later, $k$ or $\mu$ are any
symbol used to define a number is assumed to design the representation of this number
in the numeral system.

   We start by defining how we can define the number of elements of a set.

\begin{defn}\label{enum}
Let $\cal S$ be numeral system and let $E$ be a nonempty set. We say that $f$ is 
a {\bf measurement} of~$E$ in~$\cal S$ if there is a numeral in~$\cal S$ expressing 
a positive integer $\mu$, finite or infinite, with the property that $f$ is a bijection 
of $[1..\mu]$ onto~$E$. We say that a set~$E$ is {\bf measured} in~$\cal S$ if there
is a measurement of~$f$. 
\end{defn}

   When this is the case, we say that $E$ is 
{\bf measured} by $[1..\mu]$ and,
for short, that it is measured by~$\mu$. We also say that $\mu$~is the number
of elements of~$E$. We shall also denote by $\sharp E$ the number of elements of~$E$.
Note that $\mu$ is required to be 
an integer only, that it may be finite or infinite and, in the latter case, that 
it is not restricted to be bounded by~\grossone. 
Also note that when we
say that $E$ is measured, we have to be able to provide an $f$~which measures~$E$.
Moreover, this $f$~must be described, not merely assumed to exist. 
We use the words {\it measure}, {\it measured} and {\it measurement} because we
have in mind the possibility to count the number of elements beyond \grossone{} if
needed. As already mentioned about $\mu$, it must be possible to write it in some 
numeral system. This means that a measurement can be performed only if we have at our
disposal a numeral system allowing us to express the number~$\mu$ used in 
Definition~\ref{enum}.

   In Definition~\ref{enum}, we insist in the fact when a set is measured, the measurement
refers to a numeral system~$\cal S$ which has to be made explicit. As an example,
the set \hbox{$\{1,2\}$} is measured in the Pirah$\tilde{\hbox{\rm a}}$s' system
as it is measured in the standard numeral system as well as in the new one and, in
all these cases, by the identity function. However, the set \hbox{$\{1,2,3\}$}
cannot be measured in the Pirah$\tilde{\hbox{\rm a}}$s' system where number~3 cannot
be expressed. Of course, \hbox{$\{1,2,3\}$} is measured in the standard numeral system
as well as in the new one, again with the help of the identity function. This relativity
is very important, it is contained in Postulate~2: if we have a more precise language
we can see more properties.
\vskip 10pt

    Next, we consider how to define the number of elements of a subset of a set.
Starting from this point, in order to make statements easier to read, we shall not
repeat the reference to a numeral system. However, as indicated above,
we have to consider that the
word {\it measurement} always refer to such a system, as it must be possible to
express the number used to establish the
measurement in the numeral system for which the definitions or
the theorems are applied.

\begin{defn}\label{measured}
Let $A$~be a subset of the set~$E$. We say that $A$~is {\bf co-measured} in~$E$
if there are measurements $f$ and~$g$ such that $f$~measures~$A$ and~$g$ measures 
$E\backslash A$.
\end{defn}

\begin{prop}\label{num}
If $A$ is a co-measured subset of~$E$, then $E$~is measured. 
We have that $\sharp E = \sharp A + \sharp\{E\backslash A\}$.
\end{prop}

\noindent
Proof. Let $f$ be a measurement of~$A$ and let~$g$ be a measurement of $E\backslash A$.
There are positive integers~$\mu$ and~$\lambda$ such that $f$~is a bijection
from $[1..\mu]$ onto~$A$ and $g$~is a bijection from~$[1..\lambda]$ onto~$E\backslash A$.
Indeed, we define $h$ from $[1..\mu$+$\lambda]$ as follows:

\setbox110=\vtop{\leftskip 0pt\parindent 0pt\hsize=120pt
                 $f(n)$ if $n\in[1..\mu]$\vskip 5pt
                 $g(n$$-$$\mu)$ if $n\in[\mu$+$1..\mu$+$\lambda]$
                }
\ligne{\hfill
$h(n) = \left\{\vcenter{\box110}\right.$ 
\hfill}
   
\cqfd

   We can also extend the definition of the number of elements of a set in the following 
conditions.

\begin{thm}\label{bijenum}
Let $A$~be a measured set. Then $B$ has the same number of elements as~$A$ if
and only if there is a bijection from~$A$ onto~$B$. 
\end{thm}

\noindent
Proof. Assume that there is a bijection~$f$ from~$A$ onto~$B$. Let $h$~be a 
measurement of~$A$. Then $h\circ f$ is a measurement of~$B$.

   Conversely, let $f$~be a measurement from $[1..\mu]$ onto~$A$ and $g$~be an 
measurement of~$[1..\nu]$ onto~$B$. Then, as $\mu=\nu$, $g\circ f^{-1}$ is
a bijection from~$A$ onto~$B$.
\cqfd

   Now, we can order the measured sets by their number of elements.

\begin{thm}\label{inject}
Let $A$ and $B$~be two measured sets. We have that $\sharp A\leq \sharp B$ if and only if 
there is an injection from~$A$ into~$B$.
\end{thm}

\noindent
Proof. Let $f$ be a measurement from $[1..\mu]$ onto~$A$ and $g$~be a measurement from
$[1..\nu]$ onto~$B$.
Assume that $\mu\leq\nu$. Then $g\circ f^{-1}$ is an injection from~$A$ into~$B$.

Conversely. Assume that there is an injection~$h$ from~$A$ into~$B$. Assume that
$\nu<\mu$. Then $h\circ f$ is a measurement from~$[1..\nu]$ onto a proper subset $C$
of~$B$. Now, by the previous theorem, $\sharp C = \sharp B$. This is a contradiction
with Postulate~3. Consequently, as the order on integers is linear, $\mu\leq\nu$.
\cqfd 

\begin{cor}\label{bernsteinn}
Let $A$ and~$B$ be two measured sets. Then if there is an injection from~$A$ into~$B$
and if there is an injection from~$B$ into~$A$, then $\sharp A = \sharp B$.
\end{cor}

\noindent 
Proof. If the injection from~$A$ into~$B$ would not be surjective, we would obtain
a proper subset of~$B$ with as many elements as~$B$, a contradiction with Postulate~3.
\cqfd

   And so, we can see that Bernstein's theorem is true for measured sets.

   We have the following property which has no counter-part in the traditional
theory:

\begin{thm}\label{inter}
Let $A$ and~$B$ be two measured sets with $\sharp A = \sharp B$. Then,
if \hbox{$A\cap B\not=\emptyset$} and \hbox{$A\not=B$}, then 
\hbox{$A\backslash (A\cap B)$} and
\hbox{$B\backslash (A\cap B)$} are nonempty sets. Moreover, if $A\cap B$ is 
co-measured in~$A$ and in~$B$, then both \hbox{$A\backslash (A\cap B)$} and
\hbox{$B\backslash (A\cap B)$} are measured,
and \hbox{$\sharp A\backslash (A\cap B) = \sharp B\backslash (A\cap B)$}.
\end{thm}
 
\noindent
Proof. Let $A$ and~$B$ be sets satisfying the assumptions of the first sentence of the 
theorem. Let $f$~be a measurement from $[1..\mu]$ onto~$A$ and $g$~be a
measurement from $[1..\mu]$ onto~$B$, as \hbox{$\sharp A = \sharp B$}. If 
\hbox{$A\cap B\not=\emptyset$} and $A\not=B$, as we cannot have $A\subset B$ 
thanks to \hbox{$\sharp A = \sharp B$} and to Postulate~3, we have necessarily
that \hbox{$A\backslash (A\cap B) \not=\emptyset$}. Similarly, from $A\not=B$ we cannot
have $B\subset A$ and so, \hbox{$B\backslash (A\cap B) \not=\emptyset$}. 
Now, if $A\cap B$ is co-measured in both $A$ and~$B$ we have three measurements
$f_A$, $h$ and $g_B$ from $[1..\mu$$-$$\lambda]$ onto \hbox{$A\backslash (A\cap B)$},
from $[1..\lambda]$ onto $A\cap B$ and from $[1..\mu$$-$$\lambda]$ onto 
\hbox{$B\backslash (A\cap B)$} respectively. The last assertion of the theorem
follows from Theorem~\ref{inject}.
\cqfd
 
\subsection{Defining numbers}
\label{define}

Starting from this subsection  and for the restof the paper, we work in the new numeral
system defined by Yaroslav Sergeyev, see~\cite{sergeyev1,sergeyev2,sergeyev3}.
It should be appropriate to extend the frame of Definition~\ref{enum} to the
definition of elements suggested in~\cite{sergeyev3}, Subsection~5.4. 
Let us illustrate this by the example of $\displaystyle{\lfloor\sqrt{\grossone}\rfloor}$. 
This number is defined as the number of elements of the set 
\hbox{$\{x\;\vert\;x^2\leq \grossone\}$}. In~\cite{sergeyev3}, Subsection~5.4,
this is generalized from the function \hbox{$x\mapsto x^2$} to any strictly increasing 
function~$g$. When we consider a positive
finite number~$n$, we know that $n$ is the number of elements 
of the set $[1..n]$. This definition is extended to any number, including \grossone{}
and beyond.
Now, consider an initial segment~$S$ of $[1..\kappa]$, where $\kappa$ is some
infinite positive integer. This means $1\in S$ and that when $x\in S$ and $y<x$,
we have $y\in S$. If $S\not=[1..\kappa]$, there is some number~$\mu$ such
that $\mu\not\in S$.
Note that if there is a number~$\sigma$ which is the number of elements of~$S$, then
$S\subset[1..\sigma]$, by definition of the measurement, here by the identity function. 
Now, if $\sigma\not\in S$, then $S\subset[1..\sigma$$-$$1]$
which means that $S$~has at most $\sigma$$-$1 elements, a contradiction. And so
we have that $S=[1..\sigma]$. Consequently, if we consider that the number of
elements of an initial segment of $[1..\kappa]$ is defined, then any 
initial segment of $[1..\kappa]$ is of the form $[1..\sigma]$ for some
number~$\sigma\leq\kappa$. But this is an assumption so that we have to formulate
it as an axiom:

\begin{axiom}\label{segment}
Let $\kappa$~be a positive integer finite or infinite, possibly greater than
\grossone.
For each initial segment~$S$ of $[1..\kappa]$, there is a number~$\sigma\leq\kappa$ 
such that $S=[1..\sigma]$. Clearly, $\sigma$ is the {\bf number of elements}
of~$S$.
\end{axiom}

   There is here a difference between the classical theory of sets. Consider
the set $S_n$ of positive integers~$k$ which are less than~$n$. Clearly, $S_n$
is an initial segment when $n>1$. In the traditional theory, 
$\displaystyle{\reunion\limits_{n=1}^{\infty} S_n} = I\!\!N$. Using the new numeral 
system,
we get that $\displaystyle{\reunion\limits_{n\leq\grossone} S_n} = [1..\grossone$$-$$1]$.

Now, we can see that 
the way we defined $\displaystyle{\lfloor\sqrt{\grossone}\rfloor}$ 
is legitimated by
Axiom~\ref{segment}. We may also define $\displaystyle{\lfloor\sqrt{\kappa}\rfloor}$  
and many other numbers can be defined in this way as \hbox{$\lfloor\log\grossone\rfloor$}
and $\lfloor\log_b\grossone\rfloor$ for any positive finite number~$b\geq 2$. However,
note that, according to Postulate~1, the number of applications of Axiom~\ref{segment}
is finite so that in fact, we enlarge the numeral system by introducing such
notations, only finitely many times and in a finite way.

   It is important to repeat that Axiom~\ref{segment} applies to sets which are
measured and so it must be possible to express the number~$\sigma$ in the same numeral
system~$\cal S$ as the one used to express the number~$\kappa$. This means that
Axiom~\ref{segment} does not apply to any set. In particular, we cannot apply
the axiom to the set of finite positive integers. The main reason is that this
set is not well described, even in the new numeral system introduced by
Yaroslav Sergeyev. According to Postulate~2, we cannot say what are the finite positive
integers, so that we cannot speak about their set with precision. However, we
can speak of the set~$\cal F_{\cal S}$ of the expressions in~$\cal S$ of the finite 
positive integers. This set is clearly an initial segment of \hbox{$[1..\grossone]$} so 
that there is a
finite positive integer $\varphi_{\cal S}$ expressible in~$\cal S$ such that 
\hbox{${\cal F_{\cal S}} = [1..\varphi_{\cal S}]$} and such that 
\hbox{$\varphi_{\cal S}$+1} is not expressible in~$\cal S$. 

   Note that this shows us that the Peano axiom works on another plan: it says that
if $n$~is a positive integer, so is~\hbox{$n$+1}. But in stating such a property, this
axiom does not consider as relevent the possibility to express both~$n$ and
\hbox{$n$+1}. To say things in other words, 
Peano axiom does not take into account practical limitations in expressing numbers
concretely.
Now, Postulates~1 and~2 tell us that we always use a language to describe objects and that
the quality of the description depends on the expressive power of the language.
In particular, we have to take into account the limitations on writing the expression
of a positive integer. 

In any numeral system~$\cal S$, there is a maximal integer $\varphi_{\cal S}$
which can be expressed in~$\cal S$. 
This limitation
is not that surprising: in most practical programming languages, there is
a constant {\tt maxint} which denotes the greatest positive integer. The operation
\hbox{{\tt maxint}+1} cannot be performed and if your program performs such an 
operation, polite compilers inform you that there was an attempt to use a non
admissible value for the indicated type. Accordingly, the application of 
Axiom~\ref{segment} to the expressions of finite positive numbers is very natural. This 
stresses the usefulness of Axiom~\ref{segment}.

Note that Axiom~\ref{segment} is in full agreement with Postulate~3 saying that 
{\it the part is less
than the whole} is always true, whatever the sets, while this principle does not
hold for Cantor's infinite cardinals. Also note that in the traditional
ordinal theory, $\omega$ and $\omega$+1 exist but $\omega$$-$1 cannot be defined as
$\omega$ is a limit-ordinal. 

The existence of $\varphi_{\cal S}$ and the discussion about {\tt maxint} give us
the possibility to distinguish the new system from non standard analysis.
Indeed, in non standard analysis, there cannot be a maximal positive finite integer
simply because Peano axioms are there valid. Accordingly, if
$n$~is a finite integer, $n$+1 always exists in non standard analysis, even if nobody
can write it, so that there cannot be a maximal finite integer. Moreover,
if $\kappa$~is an infinite integer, it is possible, in non standard analysis, to 
construct a bijection~$\vartheta$ from \hbox{$[1..\kappa]$} onto 
\hbox{$[2..\kappa$$-$$1]$}. We define~$\vartheta$ by \hbox{$\vartheta(x)=x$+$1$}
if $x$ is a finite positive number and by \hbox{$\vartheta(x)=x$$-$$1$} if $x$~is
an infinite number. Now, in the new system, this is impossible as
\hbox{$[2..\kappa$$-$$1]$} is strictly contained into \hbox{$[1..\kappa]$}, due to
Postulate~3. We shall go back to this discussion a bit later.

As a corollary of Axiom~\ref{segment} we can state the following property:

\begin{thm}\label{upperbound}
Let $\kappa$ be a positive integer, finite or infinite, and
let $g$ be a strictly increasing function over $[1..\kappa]$. Let $\mu$ be the number
of elements of the set \hbox{$G=\{x\;\vert\;g(x) \leq \kappa\}$}. We know that
\hbox{$G=[1..\mu]$}. The number $\mu$ is also characterized as the single number~$x$
such that \hbox{$g(x)\leq\kappa < g(x$$+$$1)$}.
\end{thm}

\noindent
Proof. As $\mu\in G$, $g(\mu)\leq\kappa$. If we do not have $\kappa<g(\mu$+$1)$,
then we have \hbox{$g(\mu$+$1)\leq\kappa$} so that $\mu$+$1\in G$ and, as $g$~is 
strictly increasing, $G$ is a segment and it contains \hbox{$[1..\mu$+$1]$}, a contradiction 
with the definition
of~$\mu$. And so, $\kappa<\mu$+1. The uniqueness of $\mu$ follows from the fact
that $g$ is strictly increasing.\cqfd

Note that $g(\mu$+$1)-g(\mu)$ may be infinite: when $g(x) = x^2$, we have
that $g(\mu$+$1)-g(\mu) = 2\mu+1$ and if $\mu$ is defined by the characterization of
Theorem~\ref{upperbound} with $\kappa=\grossone$, $\mu$~is clearly infinite.

Let us remark that Axiom~\ref{segment} and Theorem~\ref{upperbound} allow us
to define a lot of numbers: this is a general paradigm related to the notion
of definability. However, concretely, we may apply them to finitely
many instances of concrete formulas only so that these new tools remain in
agreement with Postulate~1 of Sergeyev's new system.
 
   Now, Axiom~\ref{segment} allows us to define an important notion: that of the
smallest element of a set. Namely,

\begin{thm}\label{min}
Let $\kappa$~be a positive integer, finite or infinite and let $A$ be a non\-empty set 
of $[1..\kappa]$. Then, there is an integer~$\mu$ in $[1..\kappa]$ such that
$\mu\in A$ and for any $n\in A$, $\mu\leq n$.
\end{thm}

\noindent
Proof.
Let $S$ be the set of $x$ in $[1..\kappa]$ such that for any~$n\in A$ $x\leq n$.
If $1\in A$, we have the smallest element of~$A$. And so, assume that $1\not\in A$.
Clearly, $S$ is nonempty and $S$~is a segment of $[1..\kappa]$. And so, from
Axiom~\ref{segment}, there is an integer~$\nu\in [1..\kappa]$ such that
$S = [1..\nu]$. If $\nu\in A$, we are done. Now if $\nu\not\in A$, then 
for all $n\in A$, $\nu$+$1\leq n$. But then, $\nu$+$1\in S$, a contradiction with
the definition of~$\nu$. 
\cqfd

   The smallest element of~$A$ is denoted by min$\,A$.

   Theorem~\ref{min} allows us to prove a stronger version of Theorem~\ref{inter}.
We start with the following property.

\begin{thm}\label{compl}
Let $E$~be a measured set and let $A$~be a measured subset of~$E$. Then $A$~is
co-measured in~$E$.
\end{thm}

\noindent
Proof. We may assume that $A$~is a proper nonempty subset of~$E$. Let $f$~be a measurement 
of $[1..\mu]$ onto~$E$ and let $g$~be
a measurement of $[1..\lambda]$ onto~$A$. Then, we define
\hbox{$h(1) = \hbox{min}\,(E\backslash A)$} and, from this, 
\hbox{$A_1 = E\backslash(A\cup\{h(1)\})$}.
We define 
\hbox{$h(n$+$1) = \hbox{min}\,(E\backslash A_n)$}
and from that, similarly, \hbox{$A_{n+1} = E\backslash(A_n\cup\{h(n$+$1)\})$}.
Let $S$ be the set of~$n$ such that $h(n)$ and $A_n$ are defined.
It is clearly an initial segment of~$[1..\nu]$ for some $\nu\geq\mu$$-$$\lambda$. And so,
it has a greatest element~$\pi$. It is plain that $A_\pi=\emptyset$. Otherwise,
we could define $h(\pi$+$1)$ and $A_{\pi+1}$, a contradiction.
Now, as $A_\pi=\emptyset$, this proves that $\pi\geq\mu$$-$$\lambda$ as $h$~is
injective. Now, as $h([1..\pi])\subseteq (E\backslash A)$ by construction,
$\pi\leq\mu$$-$$\lambda$, so that $\pi=\mu$$-$$\lambda$ and $h$~is surjective. 
\cqfd

\begin{cor}
Let $A$ and~$B$ be two measured sets with $\sharp A = \sharp B$. Then,
if \hbox{$A\cap B\not=\emptyset$}, \hbox{$A\cap B$} is measured 
and $A\not=B$, then $A\backslash (A\cap B)$ and
$B\backslash (A\cap B)$ are measured nonempty sets. Moreover,
$\sharp A\backslash (A\cap B) = \sharp B\backslash (A\cap B)$.
\end{cor}

\noindent
Proof.  From Theorem~\ref{compl}, 
$A\cap B$ is co-measured in both~$A$ and~$B$. So that Theorem~\ref{inter} applies.
\cqfd 
\vskip 5pt
   We have another important result:

\begin{thm}\label{allmeasured}
Let $\kappa$ be a positive integer, finite or infinite. Let $A$~be a non empty set of
$[1..\kappa]$. Then $A$~is measured.
\end{thm}

\noindent
Proof. We repeat the argument of Theorem~\ref{compl}. Let $A$~be a non-empty set
of $[1..\kappa]$. Then, we know from Theorem~\ref{min} that $A$ has a smallest
element. Define \hbox{$f(1)=\hbox{\rm min}\,A$} and define 
\hbox{$A_1 = A\backslash\{f(1)\}$}. Define for any positive~$n$, finite or infinite:
\hbox{$f(n$+$1)=\hbox{\rm min}\,A_n$} and \hbox{$A_{n+1} = A_n\backslash\{f(n$+$1)\}$}.
Let $S$~be the set of $x\in[1..\kappa]$ such that $f$~is defined on $[1..x]$.
As $S$~is non empty, the above application of Theorem~\ref{min} shows us that
$1\in S$. Now, it is pain that if~$x\in S$ and $y\in[1..\kappa]$ with $y\leq x$,
then $y\in S$. So that $S$ is a non empty segment of  $[1..\kappa]$. From 
Axiom~\ref{segment}, there is an integer~$\mu\in[1..\kappa]$ such that 
\hbox{$S=[1..\mu]$}. Now, \hbox{$A_\mu=\emptyset$}, otherwise, $f(\mu$+$1)$ could be defined,
and then $\mu$+$1\in S$, a contradiction with the definition of~$\mu$. Now, $f$~is injective
by construction and, by the construction of~$f$, as 
\hbox{$A_\mu=\emptyset$}, $f$~is surjective onto~$A$. And so, $f$ is a 
measurement of~$A$. 
\cqfd

   It is important here to remind the reader that Theorem~\ref{allmeasured} deals with
sets which are clearly described only. This is why the theorem says "Let $A$~be a non
empty set of..." and not "for any non empty set of...". We have to also remark that in
most cases of a concrete set, the measurement is given with the description of the set.
Also, we remind the reader that the number used to measure a set has to be expressed
in an explicit numeral system. According to Postulate~1, we can only perform finitely
many operations on finitely many objects. Accordingly, each time we apply 
Theorem~\ref{compl}, its corollary and Theorem~\ref{allmeasured}, we can give
appropriate expressions: we use only finitely many symbols.

   As an example of a set for which we cannot immediately number its elements,
we can indicate the set of {\bf prime} integers, were a positive number greater than~1,
finite or infinite is prime whether it has two divisors exactly: 1 and itself. This set
is clearly infinite but, at the present moment, we cannot say that it is measured and, 
also, we cannot prove that it cannot be measured.
\vskip 10pt
   Now, let us consider an infinite positive integer~$\kappa$, and let us
consider the transformation $\iota\, :\, x\mapsto \kappa$+1$-$$x$. It maps $[1..\kappa]$ onto
itself and it is clearly a bijection as it is involutive. Now, it is easy to see that if
$x,y\in[1..\kappa]$, then $x<y$ if and only if $\iota(x)>\iota(y)$. This allows us
to state the following property:

\begin{thm}\label{lemax}
Let $\kappa$~be an infinite positive integer and let $A$~be a non empty subset 
of~\hbox{$[1..\kappa]$}.
Then $A$~contains an element~$x$ such that for any $y\in A$, $y\leq x$. We say that
$x$~is the greatest element of~$A$ and it is denoted by \hbox{\rm max}~$A$.
\end{thm}

\noindent
Proof. Let $\overline{A}$ be the image of~$A$ under~$\iota$. Then, as $\iota$~maps
$[1..\kappa]$ onto itself, by Theorem~\ref{min}, $\overline{A}$~has a smallest
element~$m$. Let $x=\iota(m)$. For $y\in A$, we get $\iota(y)\geq m$ and so,
$x=\iota(m)\geq \iota(\iota(y))=y$. Accordingly, $x$~is the greatest element of~$A$.
\cqfd

Now, we can define the notion of final segment:

\begin{defn}\label{final}
Let $\kappa$~be an infinite positive integer. Say that a nonempty subset $F$ 
of $[1..\kappa]$ is a {\bf final segment} if $\kappa\in F$ and, for any $x$ in~$F$ 
and any~$y\in [1..\kappa]$, from $x\leq y$, it follows that $y\in F$. 
\end{defn}

Now, it is clear from this definition that $A$~is a final segment of $[1..\kappa]$
if and only if $\overline{A}=\iota(A)$ is an initial segment of $[1..\kappa]$.
We obtain:

\begin{thm}\label{fsegment}
Let $\kappa$~be an infinite positive integer. A nonempty subset~$F$ of $[1..\kappa]$
is a final segment of $[1..\kappa]$ if and only there is an integer~$\nu\in[1..\kappa]$
such that $F=[\nu..\kappa]$.
\end{thm}

\noindent
Proof. Apply Axiom~\ref{segment} to $\overline{F}$ and then, apply again~$\iota$
as $\iota(\overline(F))=F$.
\cqfd

   Accordingly, any nonempty subset $A$ of~$[1..\kappa]$ has a smallest element and
a greatest one. This allows us to define the {\bf convex hull} of~$A$ for
any non-empty set~$A$ of $[1..\kappa]$.

\begin{defn}\label{convex}
Let $\kappa$~be an infinite positive integer. A nonempty subset~$A$ of $[1..\kappa]$
is {\bf convex} if and only if for any $x,y\in A$ with $x\leq y$, then 
$[x..y] \subset A$. If $A$~is any non-empty subset of $[1..\kappa]$,
its {\bf convex hull} is the smallest convex subset included in $[1..\kappa]$ which
contains~$A$.
\end{defn}

\begin{thm}\label{convexhull}
Let $\kappa$~be an infinite positive integer. Let $A$~be a non empty subset
of $[1..\kappa]$. Then $A$~has a convex hull which is 
$[\hbox{\rm min}\,A..\hbox{\rm max}\,A]$.
\end{thm}

\subsection{Discussion}

   We would like to discuss a few points about the results of this paper.

   First of all, remember that in Subsection~\ref{define}, we have
considered $\kappa$ as an infinite positive integer. We have mentioned after
Definition~\ref{enum} that we may consider infinite integer which are greater
than~\grossone. In Yaroslav Sergeyev's works, it was several times indicated
that there are sets whose number of elements are greater than~\grossone. Let us
give the following examples given in~\cite{sergeyevFH,sergeyevTUR,sergeyevLL}. 
The set of integers, $Z\!\!\!Z$, has $2\grossone$+1 elements. The set 
\hbox{$P= \{(a_1,a_2)\;\vert\;a_1,a_2\in I\!\!N\}$} has \grossone$^2$ elements
and the set of numeral expressions of the form $(.a_1a_2...a_{\grossone})_b$ 
with \hbox{$0\leq a_i<b$} has $b^{\grossone}$ elements. 
Now, as pointed at in~\cite{sergeyevTUR}, the word {\it sequence} is restricted
to subsets of $[1..\grossone]$ as well as the words {\it enumerate} and
{\it enumeration}. This is why in Section~\ref{yetbij} we used the words {\it measure},
{\it measured}  and {\it measurement}. Now, we did not use the term {\it measurable}
which is used in mathematics in a completely different environment. The words
{\it measure}, {\it measured}  and {\it measurement} refer to one of the
historically first physical process. The idea is to stress on the concreteness
of the notion: it evokes comparison with a yardstick to measure the length of
objects. Here we have the same idea of comparison with a yardstick: the set of
numbers up to a given one. Now, there is another reason why the word {\it measured} 
is used instead of {\it measurable}. When we say that a word is {\it measured} we 
have always to have in mind how it has been measured, {\it i.e.} we have to know at least
one way to do that, and so, we have to know at least one {\it measurement}
of the initial segment of~$[1..\kappa]$ which can be put in bijection with the
set, as well as to be able to write $\kappa$ in the numerical system we use
and the measurement explicitly describes the bijection. 

   Second, it is again a point which we have already stressed: when we speak of a set
and of an application on the set, we know a description of the set and a description
of the application. As already mentioned, the description depends on our language.
We have already indicated how the introduction of~\grossone{} allows us
to distinguish much more clearly between infinite sets than with the traditional
Cantor theory which cannot see any difference in the number of elements between
for instance $I\!\!N$ and the set of pairs of positive natural numbers.

   Now, this remark is very important. We have mentioned that Pirah$\tilde{\hbox{a}}$s
have only three numbers 1, 2 and {\it many} and that the computation rules involving
{\it many} and 1 or~2 are very similar to Cantor's rules involving $\infty$ and
finite natural number. What we have to stress here is that this difference is
very important. The new tools allow to see better, but the previous tools cannot
see what is seen by the new ones. As an example, see~\cite{sergeyevLL}, 
Pirah$\tilde{\hbox{a}}$s cannot define the set $\{1,2,3,4,5\}$. They can define the 
first two elements but the three others have no meaning for them. And so,
many problems about infinite sets which are formulated in the frame of Cantor's 
theory have a new formulation in the new numeral system and for some of them,
the problem simply vanishes. In particular, we refer the reader 
to~\cite{sergeyevFH,sergeyevRZ} for important results in this regard.

   It is now possible to make a bit more precise our discussion about the difference
between non standard analysis and the new system. We proved that for any infinite
positive number $K$, in non standard analysis, there is a bijection of \hbox{$[1..K]$}
onto \hbox{$[2..K$$-$$1]$}. The bijection~$\vartheta$ which was constructed
for that purpose cannot be defined on \hbox{$[2..K$$-$$1]$} if we consider an infinite 
positive numeral~$\kappa$ of the new system. Indeed, the representations of the finite
positive numbers which can be written in the numeral system~$\cal S$ have a maximal
number $\kappa$. Now, $\vartheta$~cannot be a measurement of \hbox{$[2..\kappa$+1]} 
as $\kappa$+1 cannot be written. In the same way, the set of infinite numbers which
can be written in~$\cal S$ is clearly a final segment and so, it has a smallest 
element $\psi_{\cal S}$. Now, \hbox{$\psi_{\cal S}$$-$$1$} cannot be written, so that
the interval \hbox{$[\varphi_{\cal S}..\psi_{\cal S}]$} contains exactly two elements
in the system $\cal S$ while in non standard analysis, if we fix $N$~as an infinite
positive integer, \hbox{$[K,P]$} contains infinitely
many integers for any finite positive number~$K$ and any infinite one~$P$ with $P\leq N$. 
Let~$\varphi$ be the function which 
maps~$x$ onto \hbox{$x$+$K$$-$1} for any finite positive
integer~$x$ in \hbox{$[1..N]$}. Then $\varphi$~is an injection from \hbox{$[1..N]$}
into \hbox{$[K,P]$}. Moreover, if we assume that \hbox{$N$$-$$P$} is a finite number,
then defining $\varphi$ on any infinite positive integer~$y$ from \hbox{$[1..N]$}
by \hbox{$y$+$P$$-$$N$}, we obtain that
$\varphi$~is a bijection from \hbox{$[1..N]$} onto \hbox{$[K,P]$}.

\section*{Conclusion}

   It seems to me that this paper stresses in a right way the importance of being
precise when looking at the number of elements of sets, especially when we wish to
compare them in this regard. We have to look at well defined sets and, when comparing
them, we also have to look at the tools on which the comparison relies.

   It seems to me that with the material given in this paper, we have more tools to
compute the number of elements of a set in the new numeral system devised by Yaroslav 
Sergeyev. I hope that this might contribute to new developments of this beautiful system.

\subsection*{Acknowledgment}

I am extremely in debt to Yaroslav Sergeyev for his attention to this work and for very
fruitful discussions, especially about the notion of relativity of our theories,
see~\cite{sergeyevFH,sergeyevRZ}. As Yaroslav always repeats, we should be fully aware 
when using mathematics that mathematics is also a living thing which evolves with 
the life of mankind.

\end{document}